# Voice Information Retrieval In Collaborative Information Seeking.

**SulaimanAdesegunKukoyi[*1], O.F.W Onifade[2], Kamorudeen A. Amuda[3]**

[1, 2, 3] **Department of Computer Science, University of Ibadan, Nigeria.**

[1*]*adesegunsulaimankukoyi@gmail.com*[2]fadowilly@yahoo.com[3]akindeleamuda@gmail.com

## ABSTRACT

*Voice information retrieval is a technique that provides Information Retrieval System with the capacity to transcribe spoken queries and use the text output for information search. CIS is a field of research that involves studying the situation, motivations, and methods for people working in a collaborative group for information seeking projects, as well as building a system for supporting such activities. Human find it easier to communicate and express ideas via speech. Existing voice search like Google and other mainstream voice search does not support collaborative search. The spoken speeches passed through the ASR for feature extraction using MFCC and HMM, Viterbi algorithm precisely for pattern matching. The result of the ASR is then passed as input into CIS System, results is then filtered have an aggregate result. The result from the simulation shows that our model was able to achieve 81.25% transcription accuracy.*

## KEYWORDS

*Information Retrieval, Collaborative Search, Collaborative Information Seeking, Automatic Speech Recognition, Feature Extraction, MFCC, Hidden Markov Model, Acoustic Model, Viterbi Algorithm.*

**INTRODUCTION**

Information Retrieval (IR) is the process of representing, storing and searching for information based on the user structured query for the purpose of knowledge gain. The process involves in Information Retrieval are as follows: Representing data, Filtering, Searching, Matching, Ranking operations and returning relevant information to the user. The main goal of Information Retrieval System (IRS) is to "finding relevant information or a document that satisfies user information needs". To achieve this goal, Information Retrieval Systems usually implement the following processes [1]:

- In indexing process the documents are represented in summarized content form.
- In filtering process all the stop words and common words are remove.
- Searching is the core process of IRS. There are various techniques for retrieving documents that match with users need.

Voice information retrieval for document (VIRD) is a technique which provides Information Retrieval System the capacity to transcribe spoken queries and use the text output for information search. The problems associated with Voice information retrieval for document (VIRD) are speech recognition errors and out-of-vocabulary (OOV) words documents [2].

Automatic Speech recognition (ASR) mechanism is a more recent approach for VIRD, it is used together with language processing for the purpose of producing transcript of spoken words which are useful for document retrieval [3].

Collaborative information seeking is a field of research that involves studying situation, motivations, and methods for people working in collaborative group for information seeking projects, as well as building system for supporting such activities.

Collaboration is often required for activities that are too complex or difficult to be dealt with by a single individual. Many situations requiring information-seeking activities also call for people to work together. Often the methods, systems, and tools that provide access to information assume that they are used only by individuals working on their tasks alone. This review points to the need to acknowledge the importance of collaboration in information seeking processes, to study models, and to develop voice activated systems that are specifically designed to enable collaborative information seeking (CIS) tasks.

Human beings find it easier to communicate and express their ideas via speech[4]. At times what we say is different from what we type.

Existing voice search like Google and other mainstream voice search do not support collaborative search. This study is therefore undertaken to introduce speech language technologies into Collaborative Information Seeking system to enhance a more natural communication between and among, the system and collaborators in need of information.

The paper is structured as follows. Section 2 explores the various literature reviews on the related work. Section 3 gives the research objectives and illustrates the methodological approach. Section 4 illustrates the results of the proposed work and Section 5 explores the conclusion and future work.

**LITERATURE REVIEW**

The most IR systems have been driven on text-based information retrieval [5]. This is because a large body of data is also available as complex unstructured data sources such text. Nowadays, text-based information retrieval is very successful because it has been studied for decades [6]. However, speech-based information retrieval is required today because of the numerous keyboard less applications such as user-friendly interface for personal computers (PCs), car navigation systems, or mobile phones [7]. To make easier for finding of relevant information on Web and the PCs, information retrieval by speech is required.

Traditionally, an information retrieval (IR) system is to provide users with those that will satisfy their information need [8]. IR allows easy access to huge amount of information (or data) [8]. It includes the use of algorithms to process a huge of unstructured or semi-structured data. The most IR systems have been driven on text-based information retrieval [5]. Today, text-based information retrieval is quite successful for IR systems because it has been studied and investigated for decades [8]. Today, with the numerous key-board-less applications, speech-based information retrieval have been studied because it al-lows the user to input a query as speech. As the result, let q be a speech query, while each document in the collection will be provided their features. Before retrieving relevant documents, the speech query must be transformed into some kind of content features such as key- words. Finally, documents can be retrieved by directly comparing them with the query [9].

Due to the advances in speech recognition technology, proper integration of IR and speech recognition has been considered and studied by many researchers [9]. In the previous study, text-based information retrieval has been investigated for decades but the study on speech-based text retrieval has just begun [6]. Unlike the text based information retrieval, text documents cannot be retrieved by directly comparing them with the speech query. It needs a process of speech-to-text in order to trans-form speech query to text before the process of text retrieval is performed.

Speech recognition is a technology that recognizes speech, allowing voice to serve as the main interface between the human and computer [10]. There is a need for a system to help the visually challenged people to provide with information like date, time, weather condition, current happenings and other general information like first search data on the internet. Speech recognition is commonly used to execute commands or write without using a keyboard, mouse or buttons [6]. In real world many tools have been introduced to address visually challenged people's problems using computer vision. But providing the same through a simple mobile with internet is the actual need [10]. There are many types of voice recognition systems in use like user dependant, independent, discrete speech recognition, continuous speech recognition, natural language recognition etc. [10]. Out of this, recognizing Natural language is difficult because recognition is done instantaneously and no other process like pattern matching is performed to identify the sounds. Such kind of system is developed that works 24 hours a day and 365 days a year. As input and output are in terms of speech.

## METHODOLOGY

**Developed VIR-CIS Model**
Figure 1 below depict the model for our voice-activated Collaborative Information Seeking System which employs MFCC and HMM for feature extraction and pattern matching respectively.

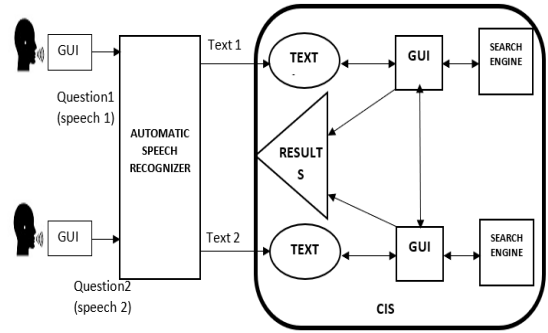

**Figure 1 Voice Information Retrieval – Collaborative Information Seeking Model**

**Procedures for VIR-CIS**

- Generation of data (Questions and Answers)
- Data Pre-Processing which entail data cleaning, MFCC feature extraction, and HMM pattern matching
- The recognized parameter is then passed into the CIS system where activities such as collaborative search, query reuse, result splitting, result merging and re-ranking took place
- Evaluation of the model

**Viterbi Algorithm for pattern matching**
The Viterbi algorithm is a dynamic programming algorithm that efficiently computes the the most likely states of the latent variables of a Hidden Markov Model (HMM), given an observed sequence of emission.

Hidden Markov Models (HMMs) model a system of discrete temporal unobserved (hidden) variables and discrete temporal observed variables. The observed and unobserved variables are related through emission probabilities.

Viterbi Algorithm pseudocode:

Function viterbi(observations of len T, state-graph of len N) returns best path:

- create a path probability matrix viterbi[N+2, T]
    - for each state s from 1 to N do: ; initialization step
        - viterbi[s, 1] ←$a_{0,s}*b_s(o_1)$
        - backpointer[s,1←0
    - for each timestep t from 2 to T do: ; recursion step
        - for each state s from 1 to N do:
            - viterbi[s,t] ←$\max_{s'=1}^{N}$viterbi[s′,t−1]∗as′,s∗bs(ot)
            - backpointer[s,t] ←$\text{argmax}_{s'=1}^{N}$viterbi[s′,t−1]∗$a_{s',s}$
    - viterbi[q_F,T] ←$\max_{s=1}^{N}$viterbi[s,T]∗$a_{s,qF}$ ; termination step
    - backpointer[q_F,T] ←$\text{argmax}_{s=1}^{N}$viterbi[s,T]∗$a_{s,qF}$; termination step
- return the backtrace path by following backpointers to states back in time from backpointer[q_F, T]

**Algorithm 1: Viterbi Pseudocode**
**source:** (Kahn, 2018)

**Collaborative Information Seeking System – CIS**
The CIS system allows collaborators or people in need of information to query individually from the data store, receive individual results, every collaborators' results is then filtered in order to have an aggregate result which is termed the combination of judgment.

This model receives the output of the Automatic Speech Recognizer as its input in order to perform the search.

Figure 2 depict the flow diagram for CIS system

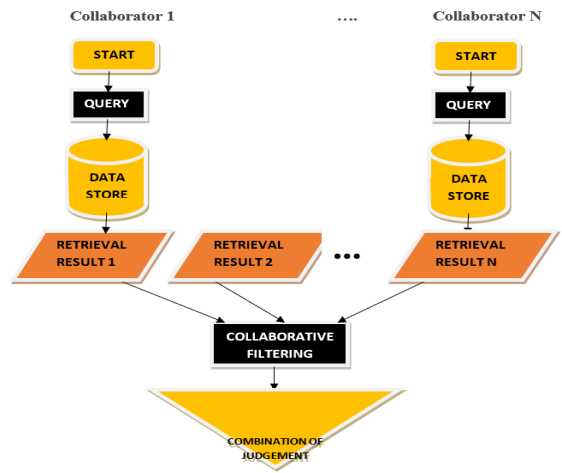

Figure 2: The flow diagram for CIS system

**Query**
This is a request for data or information from the data store which has a huge number of data. After the original speech is transformed into text with the aid of the Automatic Speech Recognizer, it is then passed into the Collaborative Information Seeking model as query in order to initiate the search process. In information retrieval the output of a search depends principally on how best the query is constructed, hence it is important that the speech is correctly transcribed in order to produce the accurate and efficient result

**Data Store**
This is the document-retrieval systems that store the entire documents, which are usually retrieved by keywords associated with the document. The text of documents is stored as data which permits full text searching, enabling retrieval on the basis of any words in the document. Pattern matching will take place in the data store in order to match the issued query with the appropriate result from the pool of data in the data store.

**Retrieval Result (1-N)**
This depicts the result obtained for nth number of collaborators. Individual collaborators have the result of his search on their graphical user interface.

**Collaborative Filtering**
Collaborative filtering, also referred to as social filtering, filters information by using the using the recommendations of other people. An automatic prediction is made merging and filtering the results of multiple users with the aid of mechanism know as relevance filter.

**Combination of Judgement**
This the final and aggregate result of Collaborative Information Seeking

# RESULTS

**Comparative analysis with existing Models in terms of transcription**
Table 1 illustrated the results of the implementation, it can be deduced that some of the previous model outperform our model in terms of transcription due to the diversity of the way we speak English as Nigerians. This means that the ASR model was not specifically meant to transcribe Nigerians mode of speaking.

**Table 1 Implementation Result**

| Author | Feature Extraction Technique | Recognition Technique | Lang | Acc. (%) |
|---|---|---|---|---|
| [11] | MFCC | HMM | Arabic | 97.99 |
| [12] | LPC | Hybrid model of Radial Basis Function and the Pattern Matching method | English | 91 |
| [12] | DWT | ANN | Malayalam | 80 |
| Our Model | MFCC | HMM (Viterbi Algorithm) | English | 81.25 |

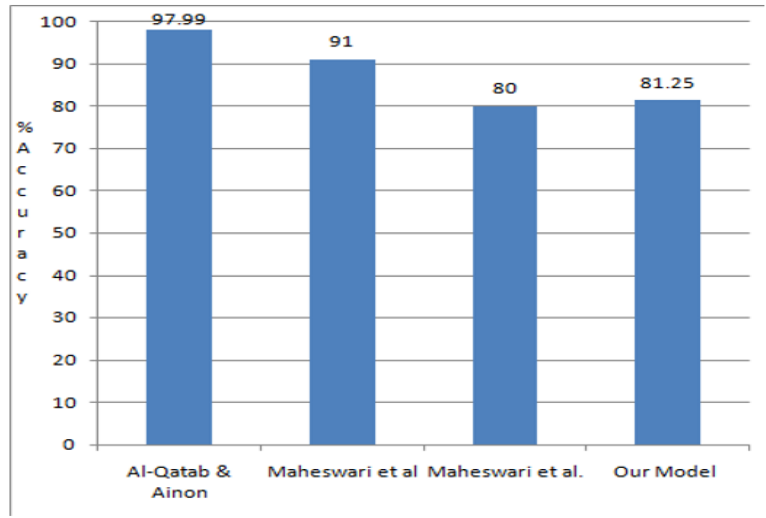

**Figure 3 Comparative analysis of implementation**

From figure 3, it can be seen that most two out of the existing model outperform our model. The former transcribed Arabic and the later transcribed English. From this we can deduced that our model transcription was hindered by the accent of the speakers because the model was not designed specifically for Nigerians.

## 3.0 CONCLUSION

The research developed a voice-activated Collaborative Information Seeking mode by using MFCC and HMM (Viterbi Algorithm) for feature extraction and pattern matching respectively. The model was evaluated and the outcome proved that Voice Information System can be incorporated into Collaborative Information Seeking model which depicts that the system is beneficial for collaborative information seeking tasks by enhancing a more natural communication flow between and among the system and collaborators in need of. The result obtained was 81.25% rate of transcription information.

However, some of the previous models outperform our model in terms of transcription due to the diversity of the way we speak English as Nigerians.

## 3.1 FUTURE WORK

This same approach could be replicated for the model designed for other Nigerian languages such as Yoruba, Igbo, Hausa etc.

**AUTHORS**

Sulaiman Adesegun Kukoyi is a graduate of University of Ibadan, Ibadan Nigeria and Tai Solarin University of Education we he bagged his MSc. and BSc(Ed) in Computer Science respectively. He is a fellow of Institute of Data Processing Management of Nigeria (IDPM), a member Computer Teachers Association of Nigeria (CTAN), Internet Society of Nigeria (ISoc), and International Association of Engineers (IAENG). His research areas covers Artificial Intelligence, Machine Learning, and Natural Language Processing (NLP).

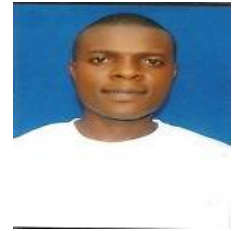

Olufade F.W Onifade is an Associate Professor of Computer Science at the University of Ibadan, Nigeria. He obtained two PhD degrees in computer science from Nancy 2 University, Nancy, France and University of Ibadan, Nigeria respectively and also published over 70 papers in both local and international referred journals and conferences. He has held several fellowships including ETTMIT and the CV Raman Fellowship for African Researchers in India. He is a member of IEEE, IAENG and CPN and his research interests include Fuzzy Learning, Information Retrieval, Biometrics and Pattern Matching.

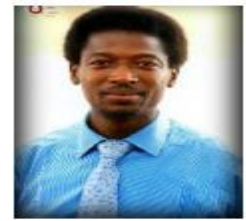

Kamorudeen AMUDA holds a BSc(Ed) in Computer Science from Adekunle Ajasin University, Nigeria and MSc. in Computer Science with a specialization in Data Mining and Analytics from the University of Ibadan, Nigeria. currently, He is a fellow Fatima Al-Fitri Predoctoral Fellowship, United State of America and also a member of Institute of Electrical and Electronics Engineers, International Association of Engineers, Association of Computing Machinery and Internet Society. His research interests are Data Mining, Machine Learning, Natural Language Processing and Artficial Intelligence

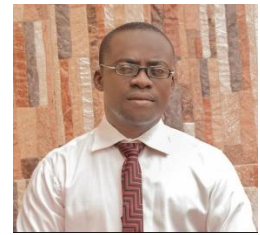